\documentclass[conference]{IEEEtran}
\usepackage{booktabs}
\usepackage{amsmath,amssymb,amsfonts}
\usepackage{algorithm}
\usepackage{algpseudocode}
\usepackage{graphicx}
\usepackage{subfig}
\usepackage{xcolor}

\begin{document}
\title{Buffer-aware Wireless Scheduling based on \\Deep Reinforcement Learning}
\author{\IEEEauthorblockN{Chen Xu\IEEEauthorrefmark{1}, Jian Wang\IEEEauthorrefmark{1}, Tianhang Yu\IEEEauthorrefmark{1}, Chuili Kong\IEEEauthorrefmark{1}, Yourui Huangfu\IEEEauthorrefmark{1}, Rong Li\IEEEauthorrefmark{1}, Yiqun Ge\IEEEauthorrefmark{2}, Jun Wang\IEEEauthorrefmark{1}}
\IEEEauthorblockA{\IEEEauthorrefmark{1}Hangzhou Research Center, Huawei Technologies, Hangzhou, China\\\IEEEauthorrefmark{2}Ottawa Research Center, Huawei Technologies, Ottawa, Canada}
Emails: \{xuchen14, wangjian23, yutianhang, kongchuili, huangfuyourui, lirongone.li, yiqun.ge,  justin.wangjun\}@huawei.com
}

\maketitle
\begin{abstract}
In this paper, the downlink packet scheduling problem for cellular networks is modeled, which jointly optimizes throughput, fairness and packet drop rate. Two genie-aided heuristic search methods are employed to explore the solution space. A deep reinforcement learning (DRL) framework with A2C algorithm is proposed for the optimization problem. Several methods have been utilized in the framework to improve the sampling and training efficiency and to adapt the algorithm to a specific scheduling problem. Numerical results show that DRL outperforms the baseline algorithm and achieves similar performance as genie-aided methods without using the future information. 
\end{abstract}

\IEEEpeerreviewmaketitle
\section{Introduction}
\label{sec:intro}
The communication field has been developed for decades with the guidance of information theory, where various advanced architectures and algorithms are proposed. However, such advanced approaches are normally designed  under traditional optimization frameworks that operate closer and closer to Shannon spectral efficiency limit. 

In recent years, deep learning (DL) has been widely applied to almost every industries and research domains, like computer vision and natural language processing. Thanks to the increasing computation power, many researchers start to resort to DL for further gain from traditional methodology. For example, instead of optimizing multiple signal processing blocks independently in traditional communication systems, an autoencoder based system is introduced to obtain a joint design architecture and finally achieves better end-to-end performance \cite{TOShea2017an}. On the other hand, conventional communication systems are characterized with rigid mathematical models that are generally linear and Gaussian-statistical to facilitate the analysis. However, neither real-world imperfections nor non-linearity can be fully represented by these linear models. To address this issue, a DL-based sequence detection algorithm is proposed for molecular communication \cite{farsad2017detection}.

Owing to the aforementioned advantages of DL, deep reinforcement learning (DRL) has been also widely employed to solve the decision making problems, turning out to be another promising technique in future communication systems \cite{luong2019applications}. The work in \cite{challita2018deep} proposes a interference-aware path planning scheme based on DRL for a network of cellular-connected unmanned aerial vehicles (UAVs). Such a scheme is able to predict the dynamics of the network, and improves the tradeoff among energy efficiency, wireless latency and caused interference. A DRL-based solution for multi-user computation offloading and resource allocation with mobile edge computation (MEC) is presented in \cite{li2018deep}, where time and energy cost is jointly minimized. Numerical results reveal that Q-learning and Deep Q network (DQN) achieve better sum cost reduction compared to the baselines. In \cite{atallah2017deep}, both safety and Quality-of-Service (QoS) are addressed with DQN in a green Vehicle-to-Infrastructure (V2I) communication scenario. While in \cite{he2017deep}, another DQN based central scheduler is proposed to obtain the optimal user selection policy in cache-enabled opportunistic interference alignment (IA) networks. A relay scheduler is introduced for the cognitive Internet of Things (CIoT) \cite{zhu2017new}, and Q-learning algorithm is designed to jointly reduce the power consumption and packet loss. In \cite{wang2019deep}, a coexistence of artificial intelligence (AI) and conventional modules are proposed, and the learning ability of DRL in cellular network scheduling problem is verified both with and without the help of expert knowledge. 

Although intensive efforts have been made on DRL based scheduling, most works consider a rather complicated scenario, which lacks a standard baseline or under some unpractical assumptions. Motivated by this, in this paper, we focus on the packet scheduling problem in cellular network to provide a more practical paradigm of AI-enabled wireless networks. Finite buffer size and maximum delay time are taken into consideration. More precisely, a DRL based scheduler is proposed,  in which a DRL agent interacts with the environment by jointly considering throughput, fairness and packet drop rate. Our main contributions are highlighted as follows.
\begin{itemize}
	\item Downlink packet scheduling in cellular network is formulated as the multi-objective optimization problem under the condition of limited transmission buffer size and queuing delay as well as link adaptation.
	\item Two genie-aided heuristic search methods are proposed to probe the performance gain in a fixed time window.
	\item Modifications of the DRL method for adapting the considering scheduling problem are presented. 
	\item Numerical results show that the DRL algorithm obtains $10.54\%$, $0.3\%$, $7.64\%$ gain over baseline in throughput, fairness and packet drop rate, respectively, achieving similar performance to the genie-aided methods without using the future information.
\end{itemize}

\section{Problem Formulation}
\label{sec:system_model}
\begin{figure}[!t]
	\centering
	\includegraphics[width=.95\columnwidth]{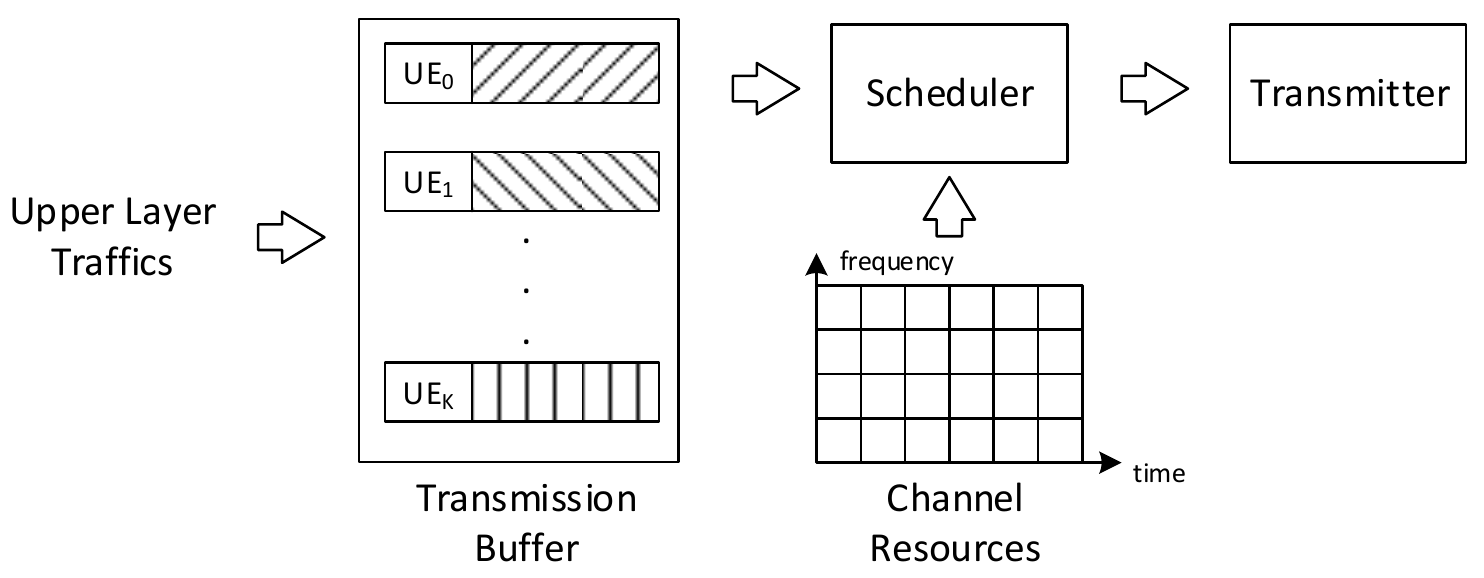}
	\caption{Transmission system model.}
	\label{fig:system_model}
\end{figure}

In a cellular network, a scheduler is of critical importance because it allocates radio resources among user equipments (UEs) while simultaneously balances between throughput and fairness \cite{capozzi2012downlink}. As shown in Fig.~\ref{fig:system_model}, $K$ active UEs in the system are waiting for being scheduled, where each traffic flow from upper layer arrives and is stored in the transmission buffer. The packet arrival is modeled as Poisson process with arrival rate $\lambda$. The scheduler allocates channel resources to UEs in each transmission time interval (TTI) according to the channel condition and buffer state. Then the head of line (HoL) packet in the corresponding buffer is sent to physical layer for transmission.

For conventional broadband systems, throughput (THP) and fairness (indicated by Jain's fairness index \cite{jain1984quantitative} (JFI)) are two key performance indicators (KPIs). Besides these, it is worthwhile to note that the packet loss due to buffer overflow and expiration is also a key factor for practical finite buffer systems. Thus,  we consider the packet drop rate (PDR) as the third KPI. The three metrics can be expressed as 
\begin{equation}
\begin{split}
&\mathrm{THP} = \sum \limits_t \sum \limits_{k \in \cal{K}} \sum \limits_{b \in \cal{B}} d_{k,b}(t)r_{k, b}(t)\\
&\mathrm{JFI} = \dfrac{\left[\sum \limits_t \sum \limits_{k \in \cal{K}} \sum \limits_{b \in \cal{B}} d_{k,b}(t)r_{k, b}(t)\right]^2}{K \sum \limits_{k \in \cal{K}} \left[\sum \limits_t  \sum \limits_{b \in \cal{B}} d_{k,b}(t)r_{k, b}(t)\right]^2}\\
&\mathrm{PDR}=\dfrac{\sum \limits_t \sum \limits_{k \in \cal{K}} (a_k(t) - s_k(t))}{\sum \limits_t \sum \limits_{k \in \cal{K}} a_k(t)}
\end{split}
\end{equation}
where $\cal K$ and $\cal B$ are set of UEs and resource block groups (RBGs), respectively. $r_{k, b}(t)$ is denoted as the achievable rate of the $k$th UE at $b$th RBG at time step $t$, $d_{k,b}(t)\in\{0,1\}$ as the scheduler decision whether $b$th RBG is allocated to UE $k$. $a_k(t)$ and $s_k(t)$ are numbers of arrived packets and transmitted packets for the $k$th UE, respectively. All the three metrics are statistically computed over a sufficient long period so that the scheduler algorithm should pay more attention to the long-term reward than short term one. 

Then the scheduling can be formulated into a multi-objective optimization problem, where the three objectives are THP maximization, JFI maximization and PDR minimization, respectively. Obviously it is NP-hard and difficult to find an optimal solution. 

\section{Genie-aided Scheduling}
\label{sec:ga}
The three objectives are intertwined to each other and hard to be optimized independently. Consequently, a trade-off among them is needed. Pareto optimization, as a classical multi-objective optimization approach, can theoretically constitute all the non-dominant tradeoffs into Pareto frontier. Then, an optimal trade-off can be selected under a typical circumstance. However, facing a long-term performance optimization problem, it is rather difficult to find out the complete Pareto frontier of the scheduling issue in practice due to the huge solution space. 

\begin{figure}[t]
	\centering
	\includegraphics[width=.9\columnwidth]{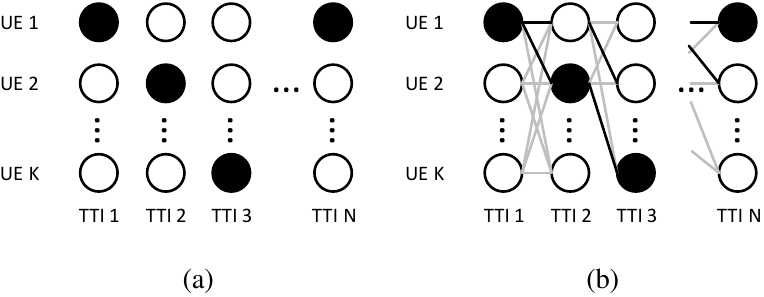}
	\caption{Search problem.}
	\label{fig:search_problem}
\end{figure}

In order to explore the gain space of the scheduling problem, we introduce two genie-aided methods based on Pareto optimization in this section. Genie-aided methods assume that the scheduler is given the information that it would not access to in practice. For example, the scheduler knows all the information in the scheduling window of $N$ TTIs. More precisely, we assume that the scheduler has obtained the channel state information (CSI) and the packets arrival of future $N$-TTI duration, then the single-RBG scheduling problem over $N$ TTIs for $K$ UEs is equivalent to a search problem that is to find one sequence of actions among $K^N$ candidate action groups, as shown in Fig.~\ref{fig:search_problem}(a) (Dark circles denote the scheduled UEs). Obviously, it is computationally forbidden to achieve the optimal solution over this immense search space. Thus, here we only present two heuristic algorithms instead, i.e., genetic algorithm (GA) and Pareto list algorithm (PLA).

\subsection{Genetic Algorithm}

Genetic algorithm (GA) are well-suited for search problems involving several, often conflicting objectives. The main goal is to obtain multiple good action groups that have high objective values and at the meantime to maintain the diversity. We adopt the nondominated sorting genetic algorithm II (NSGA-II) approach, which uses a fast nondominated sorting procedure, a diversity-preserving operator, and an elitist-preserving approach (details please refer \cite{deb2002fast}).

\textbf{Fast nondominated sorting procedure:} For each solution $p$, two entities are calculated: 1) $n_p$, the number of solutions which dominate the solution $p$; 2) $S_p$, a set of solutions that the solution $p$ dominates. Based on the values of $n_p$ and $S_p$, the population is sorted into different nondomination levels.

\textbf{Diversity-preserving operator:} For the members in the same nondomination level, we use the diversity-preserving operator which consists a fast crowded distance estimator and a simple crowded comparison procedure to guide the selection toward a uniformly spread-out Pareto-optimal front.

\textbf{Elitist-preserving approach:} Elitism helps in achieving better convergence and performance. At each generation, a combined population from $P$ parents and $P$ offsprings is sorted according to both the nondomination levels and crowded distance, and then the first $P$ population members are chosen sequentially.

For GA, the length of chromosome is $Q$, and the variable of each gene lies in the set of $\left\{1,2,\ldots,K\right\}$. The crossover and mutation operators are conducted to introduce the possibilities of generating new and better action groups.

\subsection{Pareto List Algorithm}
The Pareto list algorithm executes path expanding, sorting and pruning TTI by TTI, constraining the complexity to a fixed level, i.e., the maximum number of the list $L$. 

Firstly, at each TTI, a path is expanded according to the number of active UEs. Note that different paths may lead to different active UEs and be expanded in different ways, because paths affect the states, such as UE buffers. After expansion, each path records throughput, fairness, packet drop rate of all the UEs till the $n$th TTI.

When the number of the paths exceeds $L$, path sorting and pruning is needed to limit the path number to $L$. As shown in Fig.~\ref{fig:search_problem}(b), the gray lines mean pruning and the dark lines represent the preserved paths. Unlike the conventional list-based algorithm in which the path metric is a scalar, the path sorting and pruning in the scheduling issue is a multi-objective problem and can be handled in a Pareto-based way. In order to guarantee the global convergence and avoid being trapped in a local optimum, the optimality and diversity of the preserved paths are needed. Inspired by the NSGA-II algorithm, we propose a modified sorting method which can further improve the path diversity. In the scheduling issue, large number of paths may result in the same states, gathering at one point. Therefore, despite of the non-dominated sorting and crowded distance sorting, the paths with the same states are removed in advance. 

After the path expansion, sorting and pruning of the last TTI have been executed, a final scheduling path is selected from the path list according to the preference or the typical circumstance to satisfy the system requirements. 

\section{Deep Reinforcement Learning based Scheduling}
\label{sec:drl}
\subsection{Markov Decision Process}
In wireless networks, most decision making problems can be modeled as Markov decision process (MDP). An MDP is typically defined by a tuple $\left ({\cal{S}}, {\cal{A}}, P, r \right)$, $\cal{S}$ is the set of states, where $\cal{A}$ is the set of actions, $P(s'|s, a)$ is the transition probability from state $s$ to $s'$ due to action $a$, and $r$ is the immediate reward when transition happens.

To tackle the scheduling problem with DRL method, we first define state, action and reward as follows:

\textbf{State}. The input state contains all UEs' observations. The estimated instantaneous rate, averaged rate, the spare space in the buffer and the waiting time of the HoL packet are concatenated to form the observation of each UE, since the agent should have some knowledge of the buffer state to be buffer-aware.

\textbf{Action}. The action set consists of $K$ one-hot code, indicating which UE is selected for each transmission. Note that we reuse the same policy network for different RBGs to avoid an exponential increase of action space which incurs significant training costs and probably unconvergence.

\textbf{Reward}. As throughput, fairness and packet drop rate are three KPIs that we concern about, a straight forward reward function is defined in the form of linear weighted sum
\begin{equation}
r = \alpha thp + \beta jfi - \delta pdr
\end{equation}
where $thp$ and $jfi$ are total throughput and JFI per TTI. $pdr$ is the total number of dropped packets that are normalized with $K$ at the same interval. $\alpha$, $\beta$ and $\delta$ are the weighting factors. Although linear scalarization of a multi-objective problem may lead to non-convex Pareto frontier, we find it still efficient to obtain a satisfying result in our training framework.

\subsection{Deep Reinforcement Learning based Method}
The MDP is usually solved by classical dynamic programming algorithms, e.g., value iteration or policy iteration if the state transition probability is perfectly known. When the problem becomes complicated and large-scale, model-free DRL methods are alternatives for handling such situation. At each time step $t$, the DRL agent observes state $s_t$ from environment, makes decision $a_t$ then receives reward $r_t$ and next state $s_{t+1}$. The goal of DRL is to find a policy $\pi\left( a|s\right)$ through such interaction with environment that maximizes the accumulated (discounted) reward.

Advantage actor-critic (A2C) algorithm is employed to solve the scheduling problem in this paper. The actor-critic is essentially a policy-based DRL algorithm which directly parameterizes the policy as $\pi_\theta(a|s)$. The parameters $\theta$ can be updated by gradient ascent on the expected return $R$:
\begin{equation}
\label{eq:gradient}
g=\nabla_\theta \mathbb{E}\left[R\right] = \nabla_\theta \mathbb{E}\left[\sum_{t=0}^{\infty} \gamma ^ t r_t \right]
\end{equation}
where $\gamma$ is the discount factor which determines the importance of the future reward. The gradient in \eqref{eq:gradient} can be further represented as \cite{schulman2015high}:
\begin{equation}
\label{eq:gae}
g = \mathbb{E}\left[ \sum_{t=0}^{\infty}\Psi_t \nabla_\theta \log \pi_\theta \left(a_t | s_t\right)  \right]
\end{equation}

\begin{equation}
\label{eq:adv}
\begin{split}
A^\pi\left(s_t, a_t\right) = Q^\pi\left(s_t,a_t\right)-V^\pi\left(s_t\right)\\
Q^\pi\left(s_t,a_t\right) = {\mathbb{E}}_{\substack{s_{t+1}:\infty\\a_{t+1}:\infty}} \left[\sum_{l=0}^{\infty}\gamma^l r_{t+l}\right]\\
V^\pi\left(s_t\right) = {\mathbb{E}}_{\substack{s_{t+1}:\infty\\a_t:\infty}} \left[\sum_{l=0}^{\infty}\gamma^l r_{t+l}\right]\\
\end{split}
\end{equation}

The advantage function in \eqref{eq:adv} shows whether the action $a$ is better than the average performance of the current policy $\pi_\theta(a|s)$. That means if it is true, the gradient update should increase the probability of this action.

The estimated gradient will have much lower variance if $\Psi_t = A^\pi \left(s_t, a_t\right)$, however, the advantage function needs to be estimated by another parameterized value function $V^\pi_\phi\left(s_t\right)$:
\begin{equation}
\label{eq:adv_est}
A^\pi \left(s_t, a_t\right) = \mathbb{E}_{s_{t+1}} \left[r_t + \gamma V^\pi\left(s_{t+1}\right) - V^\pi\left(s_t\right)\right]
\end{equation}

\begin{figure}[!t]
	\centering
	\includegraphics[width=.9\columnwidth]{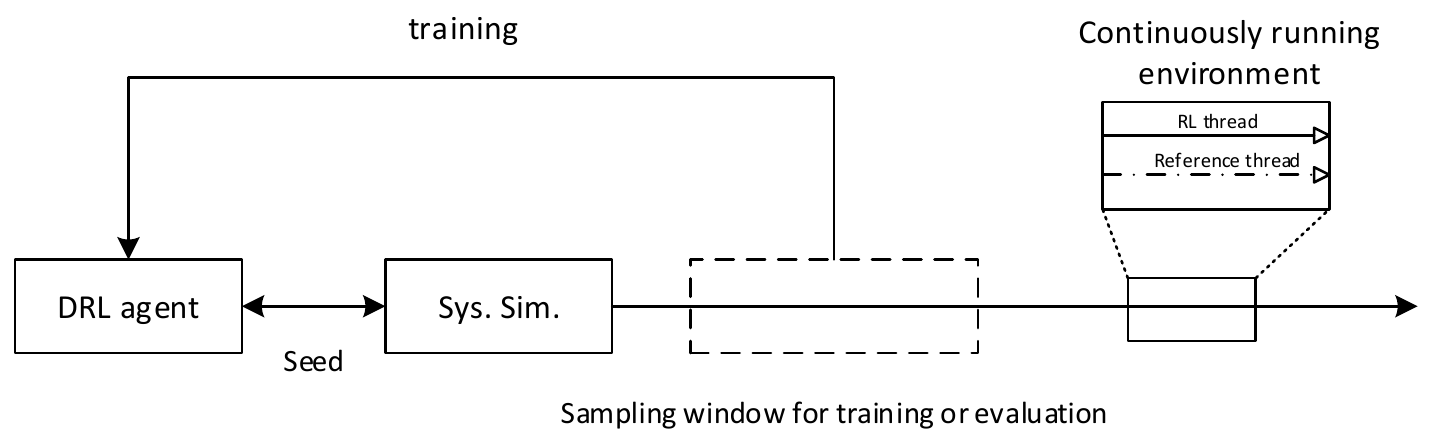}
	\caption{Framework of the DRL method.}
	\label{fig:framework}
\end{figure}

The system-level simulator is shown in Fig.~\ref{fig:framework}. DRL agent learns from the training data which is generated by interacting with the simulator. Training and evaluation data are obtained by setting a certain sampling window in which both DRL and reference performance are collected at the same environment state for a fair comparison. To make the original A2C algorithm applicable to our scheduling problem, we make the following efforts:

\textbf{$\boldsymbol{n}$-step return.} The estimation of the advantage function in \eqref{eq:adv_est} is called one-step temporal difference (TD). In fact, we implement a more general version $n$-step TD in our framework. Instead of sampling from $s_t$ to $s_{t+1}$, more actions are taken to get $n$ rewards along the trajectory. This helps improve the advantage estimation by averaging out some variance during gradient updates and leads to a stabler training.

\begin{equation}
\label{eq:nstep}
\begin{split}
A^\pi \left(s_t, a_t\right) = \mathbb{E}_{s_{t+1}\cdots s_{t+n}} [&r_t + \gamma r_{t+1} + \gamma^2 r_{t+2} +\cdots \\
&+ \gamma ^ n V^\pi\left(s_{t+n}\right)- V^\pi\left(s_t\right)]
\end{split}
\end{equation}

\textbf{Entropy regularization.} Although there are several environments providing uncorrelated experiences at the same time, the policy still easily converges to a deterministic local optimal. Here, an entropy regularization in \eqref{eq:entropy} for policy network is employed to enhance the exploring ability of the DRL agent, since the exploring only comes from the sampling on the policy distribution. 

\begin{equation}
\label{eq:entropy}
H=-\sum\limits_a \pi_\theta(a|s)\log\pi_\theta(a|s)
\end{equation}

\begin{figure}[!t]
	\centering
	\includegraphics[width=.9\columnwidth]{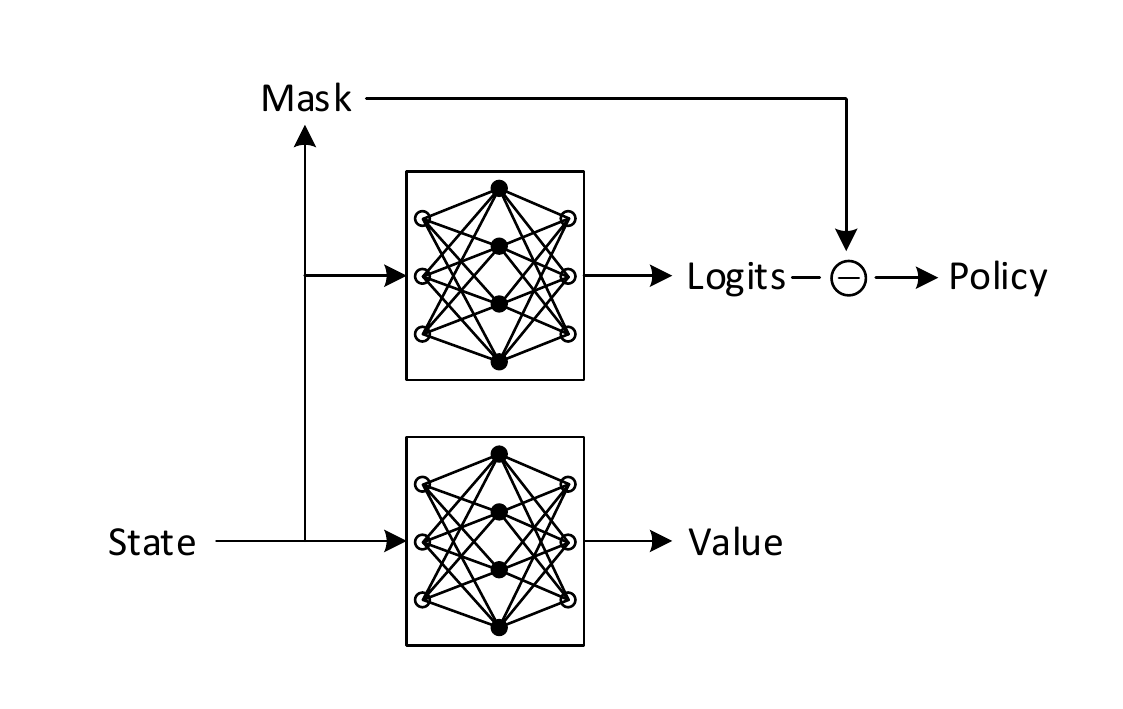}
	\caption{Neural network structure.}
	\label{fig:nn}
\end{figure}

\textbf{Action masking.} The policy and value network in our A2C network are both fully connected networks. Special structure is applied on the policy network to handle the inactive UE problem introduced by the non-full buffer scenario as in Fig.~\ref{fig:nn}. A mask is generated by the states, and prohibits policy from choosing the inactive UE(s), e.g., by subtracting a large value from the corresponding logits, then a softmax activation function is adopted to output the probability distribution. 

\textbf{Multiple-RBG scheduling.} Multiple-RBG scheduling is also considered on the basis of single RBG in our scheme. Assume that there are $B$ RBGs, the maximum dimension of the action space is $K^B$. An iterative reuse of the policy network among the RBGs is adopted in this paper to deal with the dimension curse problem. A policy network for $K$-UE scheduling is designed, where the state in the same time step will be adjusted by the possible influence of the action that already carried out.

To sum up, the full algorithm is described in Alg.~\ref{alg:a2c}.
\begin{algorithm}[!t]
	\caption{A2C algorithm}
	\label{alg:a2c}
	\begin{algorithmic}
		\State Initialize all environments
		\State Initialize actor network $\pi_\theta$ and critic network $V_\phi$
		\State Initialize experience buffer $E$
		\For {iteration = 1, $M$}
		\State \Call{Sample\_Batch}{$n, \pi_\theta$}
		\State Update discounted reward $r_i$ for $i$th experience
		\State Policy objective $J_\theta = \sum_{i} \left(r_i - V_\phi(s_i)\right) \log \pi_\theta (a_i | s_i) $
		\State Entropy term $H_\theta = - \sum_{i}\pi_\theta(a_i|s_i) \log \pi_\theta(a_i|s_i)$
		\State MSE of value $L_\phi = \sum_{i} \left(r_i - V_\phi(s_i) \right)^2$
		\State SGD with $G = - \left(\nabla_\theta J_\theta + \lambda_e \nabla_\theta H_\theta \right) + \lambda_v\nabla_\phi L_\phi$
		\EndFor
		
		\Function{Sample\_Batch}{$n, \pi_\theta$}
		\State Clear $E$
		\For {t = 1, $n$}
		\State Observe $s_t$
		\State Choose action $a_t \sim \pi_\theta(s_t)$
		\State Take action $a_t$, observe $s_{t+1}$ and $r_t$
		\State Store ($s_t, a_t, r_t, s_{t+1}$) into $E$
		\EndFor
		\EndFunction
	\end{algorithmic}
\end{algorithm}

\section{Numerical Results and Discussions}
\label{sec:exp}
\begin{figure}[!t]
	\centering
	\includegraphics[width=.9\columnwidth]{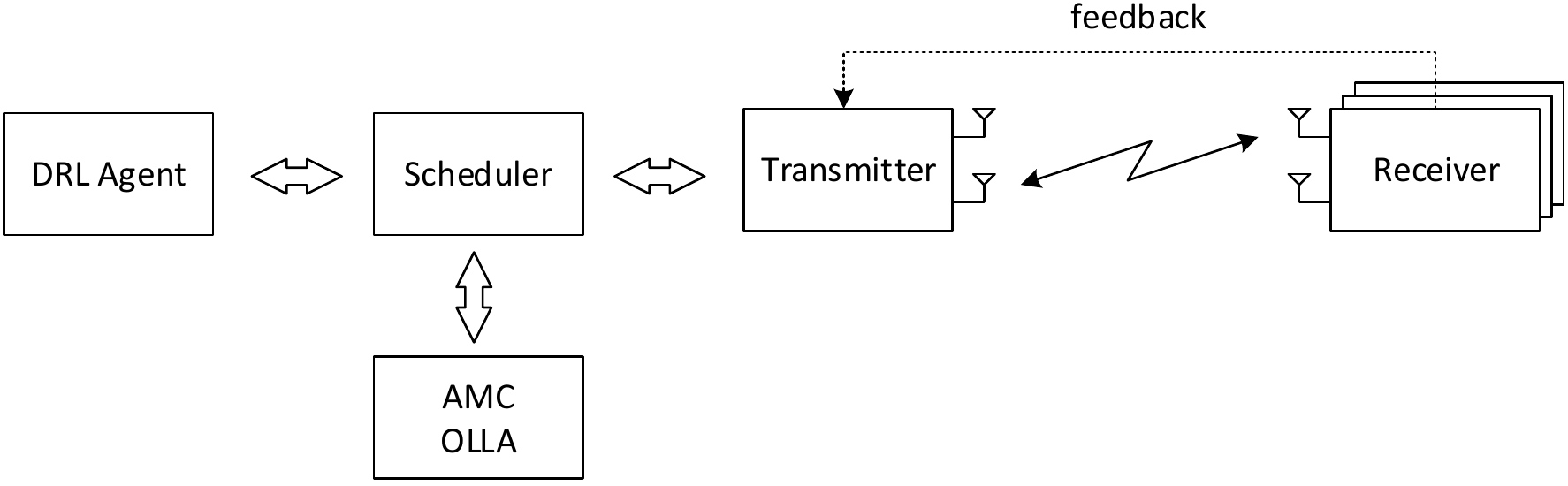}
	\caption{System-level simulator.}
	\label{fig:sls}
\end{figure}

An LTE based system-level simulator is used to evaluate the proposed DRL method, as shown in Fig.~\ref{fig:sls}. The scheduler allocates physical resources for packet delivery to UEs periodically. The feedback mechanism is established between transmitter and receiver so that link adaptation techniques can be applied at the transmitter. Herein, LTE standard adaptive modulation and coding (AMC) function is adopted where modulation and coding schemes (MCSs) are chosen in order to achieve a target block error ratio (BLER). And outer loop link adaptation (OLLA) is also realized to provide fixed step compensations to the feedback imperfection based on ACK/NACK signaling that UEs report. An DRL module is integrated in the simulator to make a better decision. Proportional fairness (PF) scheduling \cite{tse2001multiuser} is baseline algorithm in the system.

\begin{equation}
\label{eq:pf}
i = \mathop {\arg \max }\limits_{k \in \cal{K}} {{I_k } \over {T_k }}
\end{equation}
where $I_k$ is the estimated instantaneous rate and $T_k$ is the exponential moving average throughput of $k$th user.

To improve the sampling efficiency of the on-policy A2C algorithm, 56 simulators for a same DRL agent are launched simultaneously for a quickly and adequately exploring in state and action space. The UE deployment and random seed for each simulator are differentiated in order to decrease the data correlation in a batch which is lethal for DRL. In addition, averaging among different random seeded simulators also increases the generalization and are more reasonable in performance evaluation. 

\begin{figure}[!t]
	\centering
	\includegraphics[width=\columnwidth]{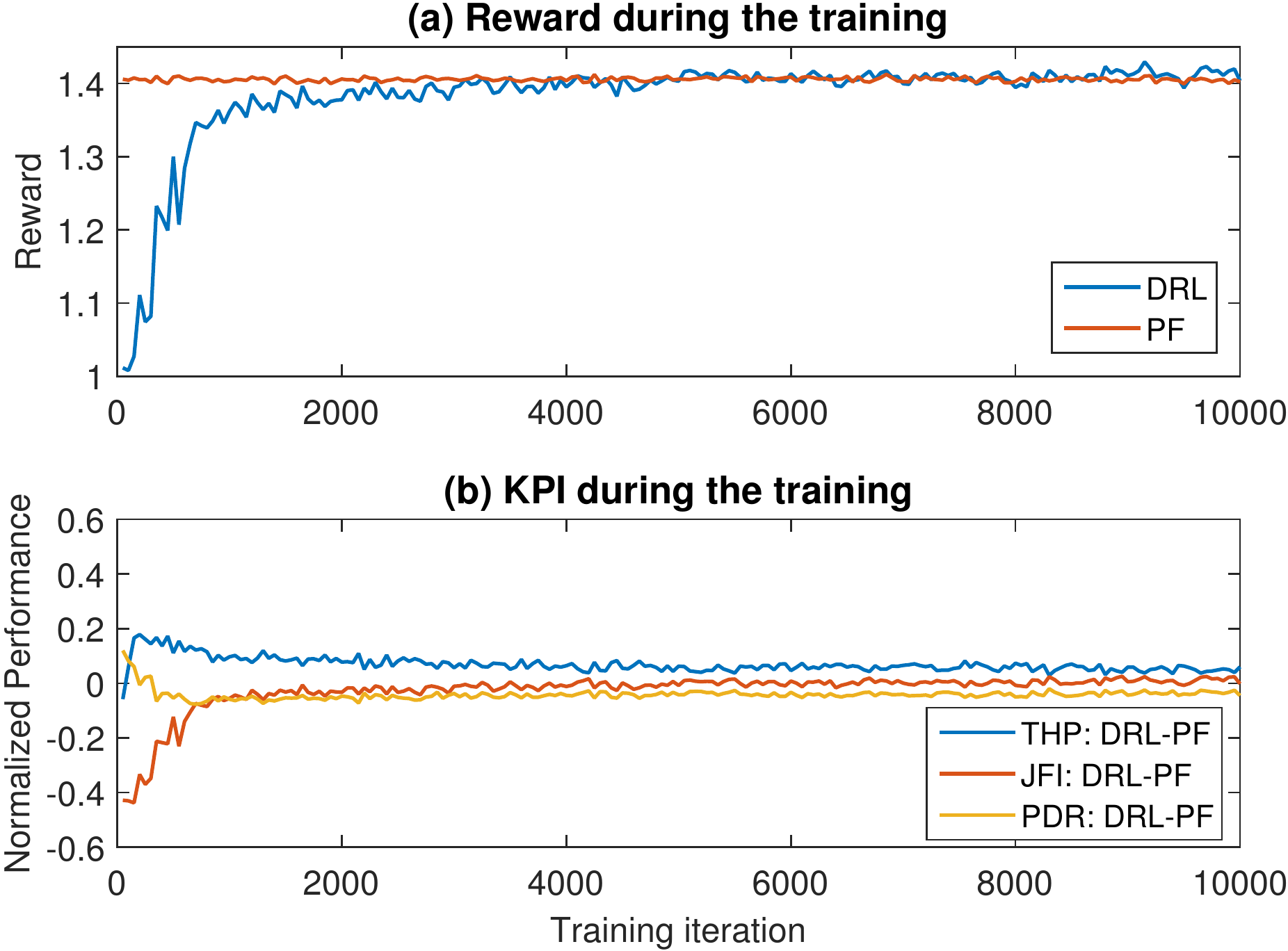}
	\caption{Training of the DRL.}
		\label{fig:training}
\end{figure}

The simulation is conducted with $K=5$ UEs and scheduling window $N=500$ TTIs, the maximum number of RBGs is $B=10$. The neural networks (NNs) used in the DRL agent are fully connected ones with 2 hidden layers, each of which contains 640 neurons. ReLU function is used as the activation function for all the hidden layers. The policy network outputs decision with softmax and linear activation is used for value function.

Fig.~\ref{fig:training}(a) shows the reward that DRL agent obtains during the training iteration. Also, we plot the reward value of PF algorithm for reference. After $\sim$5000 updates, the DRL obtains the same reward value as the PF algorithm and converges. The learning rate is further decayed at iteration 5000 so that the DRL finally achieves a larger reward than PF algorithm. The variation of the KPI values during the same training is recorded in Fig.~\ref{fig:training}(b). The performance is evaluated every 50 updates and 
the normalized performance gaps between DRL and PF algorithm is elaborated. It is interesting to see that the agent quickly learns a scheduling policy similar to MAXC/I, then slowly converges to the policy that outperforms the PF algorithm in all three KPIs.

\begin{figure}[!t]
	\centering
	\includegraphics[width=.9\columnwidth]{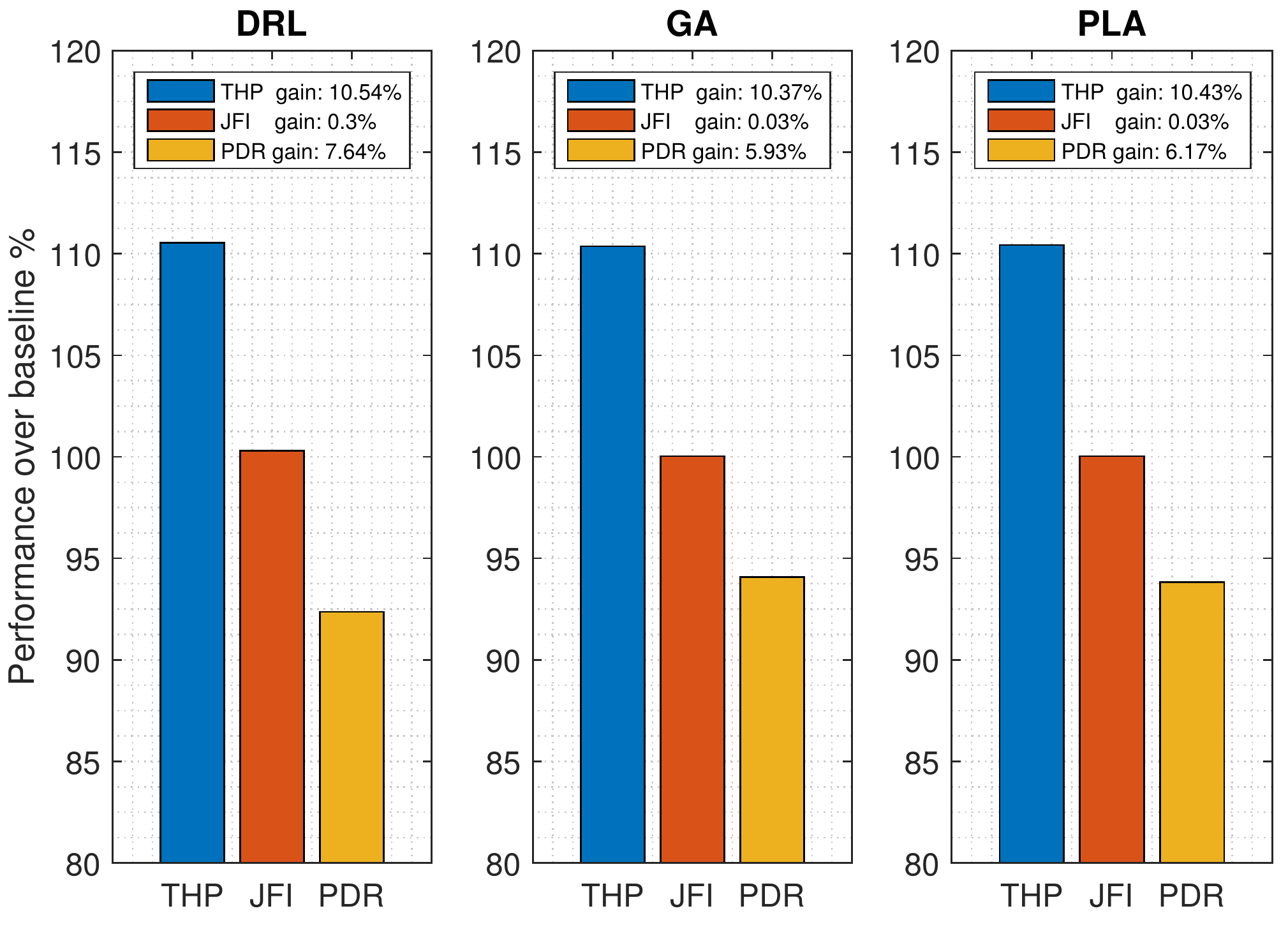}
	\caption{Performance metrics of DRL and genie-aided methods.}
	\label{fig:singlerbg}
\end{figure}

\begin{figure}[!t]
	\centering
	\includegraphics[width=.9\columnwidth]{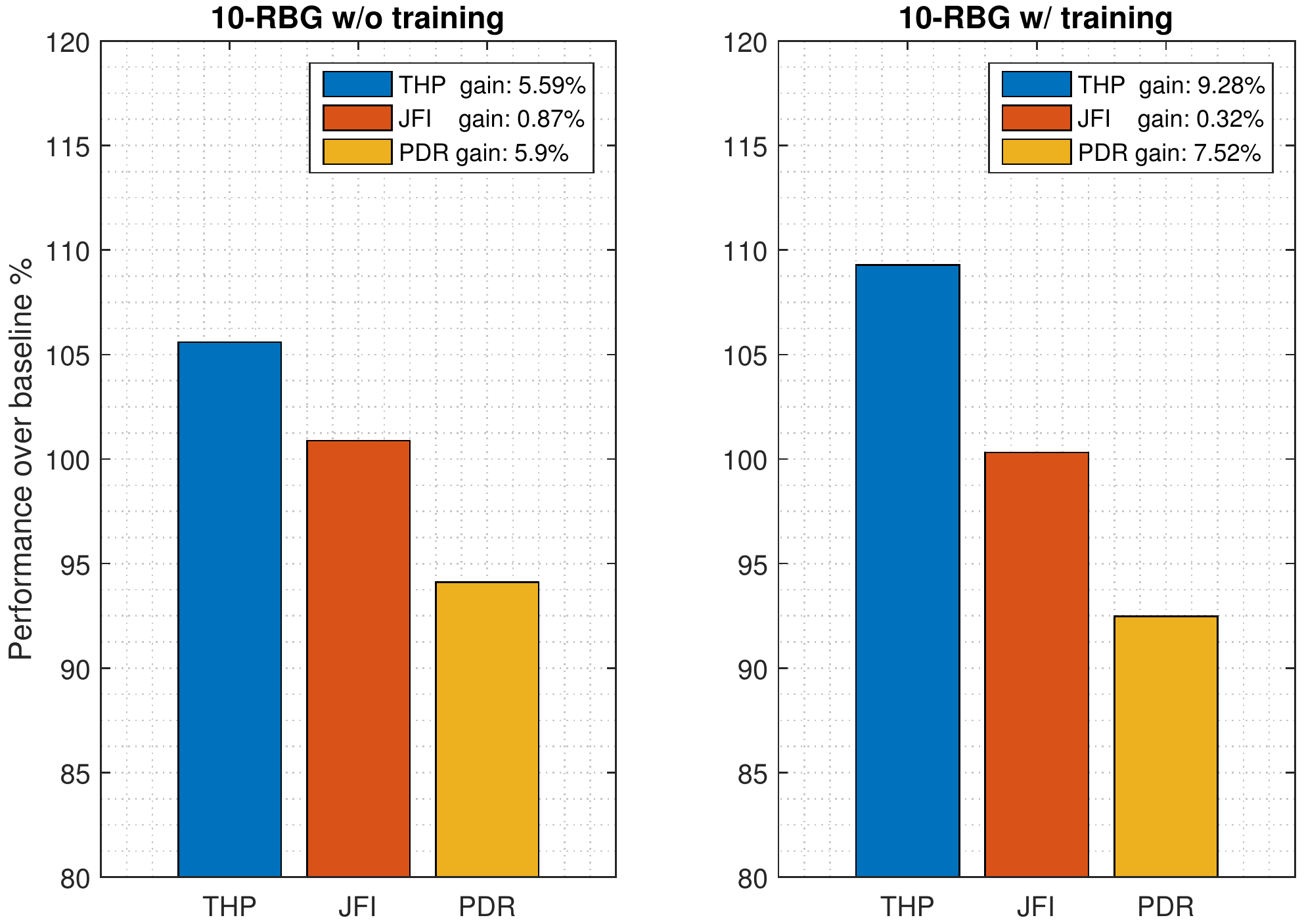}
	\caption{Performance of DRL for multiple RBGs.}
	\label{fig:multirbg}
\end{figure}

After training, we fix the parameters of the NN model and run a performance evaluation. The average performances over baseline in 20000 TTIs of 56 UE deployments are shown in Fig.~\ref{fig:singlerbg} and Fig.~\ref{fig:multirbg}. For GA, we use simulated binary crossover operator and polynomial mutation with the crossover probability of $p_c = 0.95$ and a mutation probability of $p_m = 0.05$. Also, the distribution indexed for crossover and mutation operators are $\eta_c = 5$ and $\eta_m = 20$, respectively. We set the population size as $3000$ and $10000$ generations to conduct the GA. For PLA, the maximum list number of PLA is $L=2000$. Both GA and PLA considers the single RBG case due to the complexity.

\begin{figure}[!t]
	\centering
	\includegraphics[width=.9\columnwidth]{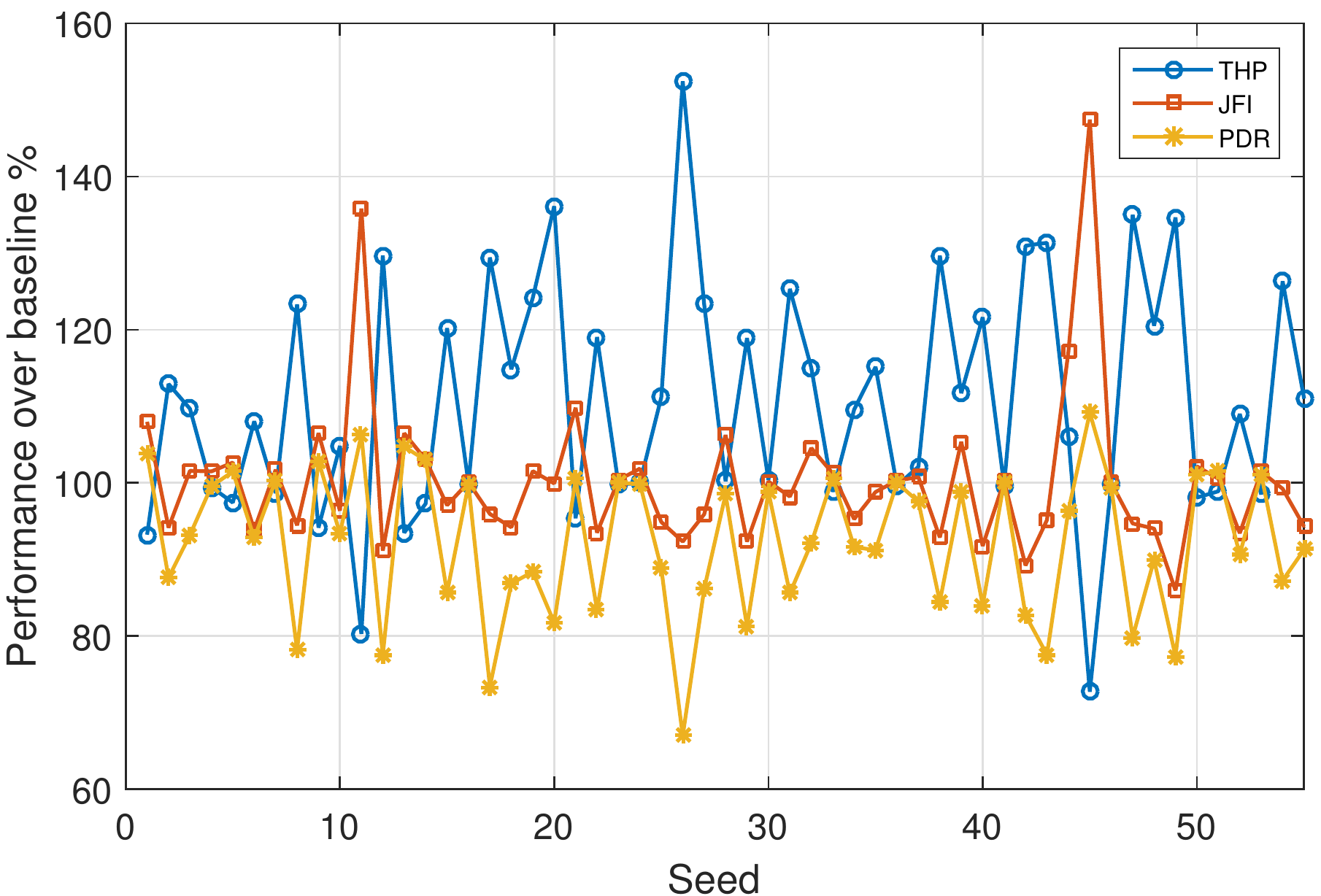}
	\caption{Performance metrics for DRL of each seed.}
	\label{fig:seed}
\end{figure}

For single RBG as in Fig.~\ref{fig:singlerbg}, we can see that the performances of GA and PLA are similar in our configuration. DRL obtains nearly the same throughput, but slightly better JFI and PDR. It should be noted that DRL algorithm only has one policy for all UE deployments and has no future information when making decisions, which is very different from genie-aided method. The gain of DRL merely comes from the learning ability by exploring and getting rewards. Fig.~\ref{fig:seed} plots the performance of DRL in all 56 deployments, where we can see that the THP and PDR gain are obvious among all seeds while JFI keeps almost the same to the baseline. We argue that some THP vales, e.g., seed 11 and seed 45, are not failures because they have large JFI, which means they are still somewhere near the Pareto frontier. We believe that in real world, some deployment-specific KPI weightings will help them fast converges to the required performance.  

As for multiple RBG scheduling, two methods have been tried: a) transfer learning, i.e., the NNs trained for single RBG of 1.4 MHz bandwidth is directly reused in the 10 RBGs system with 20 MHz bandwidth, without any further retraining. b) training the new model that fit for 10-RBG system. Both results are illustrated in Fig.~\ref{fig:multirbg}, we find that the model in first method exhibits great generalization capability in transfering to a system with more resources. The performance is mildly degraded but still better than the baseline. The re-training method further exploits the learning ability of the agent and achieves a similar performance to single RBG.

\section{Conclusion and future work}
\label{sec:conclusion}
In this paper, we propose DRL method to solve the scheduling problem in cellular networks. The practical scheduling issue is modeled as a multi-objective optimization problem consisting of long-term throughput maximization, fairness maximization and packet drop rate minimization. Two genie-aided methods are employed to probe the performance gain space. Then a modified A2C algorithm is proposed to solve the considered scheduling problem. The results show that the DRL can outperform baseline PF algorithm and achieve similar performance to the genie-aided methods without using the future information. In the future, multi-agent reinforcement learning (MARL) structure is considered  to further improve the spectral efficiency, e.g., by providing the ability of intelligent inter-cell interference cancellation.

\bibliographystyle{IEEEtran}
\bibliography{reference}

\end{document}